\documentclass[reprint,10pt,aps,prl]{revtex4-2}
\pdfoutput=1

\usepackage{amsmath}
\usepackage{amssymb}
\usepackage{color}
\usepackage{graphicx}
\usepackage[hidelinks]{hyperref}
\usepackage[utf8]{inputenc}
\usepackage[english]{babel}
\usepackage{stix}
\usepackage{dsfont}
\usepackage{xspace}
\usepackage[normalem]{ulem}
\usepackage{tabularx}
\usepackage{xcolor}
\usepackage{csquotes}
\usepackage{mathtools}

\newcolumntype{Y}{>{\centering\arraybackslash}X}
\newcolumntype{L}[1]{>{\raggedright\let\newline\\\arraybackslash\hspace{0pt}}m{#1}}
\newcolumntype{C}[1]{>{\centering\let\newline\\\arraybackslash\hspace{0pt}}m{#1}}
\newcolumntype{R}[1]{>{\raggedleft\let\newline\\\arraybackslash\hspace{0pt}}m{#1}}

\newcommand{\ie}{i.\,e.\@\xspace}

\newcommand{\Lim}[1]{\raisebox{0.5ex}{\scalebox{0.8}{$\displaystyle \lim_{#1}\;$}}}
\newcommand*{\white}{\textcolor{white}}

\newcommand*{\Vghost}{\white{$a^{M^{M}}$}}

\begin{document}
	\title{Breitenlohner-Freedman bound on hyperbolic tilings}
	\author{Pablo Basteiro}
	\author{Felix Dusel}
	\author{Johanna Erdmenger}
	\email{erdmenger@physik.uni-wuerzburg.de}
	\author{Dietmar Herdt}
	\author{Haye Hinrichsen}
	\author{Ren\'e Meyer}
	\author{Manuel Schrauth}
	\affiliation{Institute for Theoretical Physics and Astrophysics and W\"urzburg-Dresden
		Excellence Cluster ct.qmat, Julius Maximilians University W\"urzburg, Am Hubland, 97074 Würzburg, Germany}
	
	\begin{abstract}  
		We establish how the Breitenlohner-Freedman (BF) bound is realized on tilings of two-dimensional Euclidean Anti-de Sitter space. For the continuum, the BF bound states that on Anti-de Sitter spaces, fluctuation modes remain
		stable for small negative mass-squared $m^2$. This follows from a real and positive total energy of the gravitational system. For finite cutoff $\varepsilon$, we solve the Klein-Gordon equation numerically on regular hyperbolic tilings. When $\varepsilon\to0$, we find that the continuum BF bound is approached in a manner independent of the tiling. We confirm these results via simulations of a hyperbolic electric circuit. Moreover, we propose a novel circuit including active elements that allows to further scan values of $m^2$ above the BF bound.
	\end{abstract}
	
	\maketitle

	\paragraph{Introduction:}\label{sec:Intro}
	
	The AdS/CFT correspondence \cite{Maldacena:1997re,Gubser:1998bc,Witten:1998qj}, also known as holography, maps gravitational theories in \mbox{$(d+1)$}-dimensional hyperbolic Anti-de Sitter (AdS) spacetimes to strongly coupled conformal field theories (CFTs) without gravity in $d$ dimensions, defined on the AdS boundary. The AdS/CFT duality provides a precise map between CFT operators and AdS gravity fields, which is of great significance both for fundamental aspects of quantum gravity \cite{Susskind:1994vu}   and for applications to strongly correlated condensed matter systems \cite{Hartnoll:2009sz}.
	
\begin{figure}[t]
		\centering
		\includegraphics[width=0.73\columnwidth]{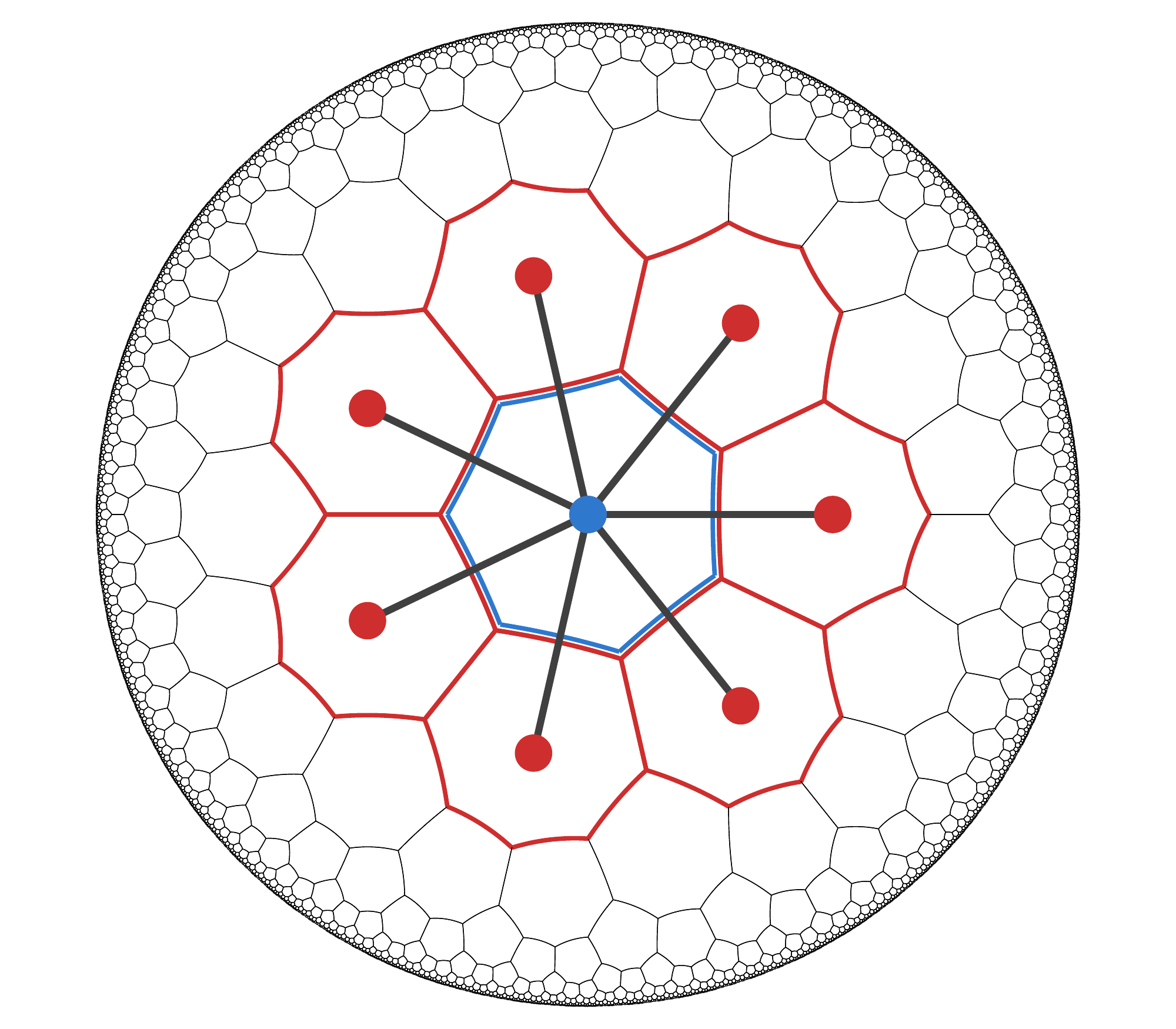}
		\caption{Hyperbolic \{7,3\} tiling in the Poincare disk representation. For the central node, the stencil of the discretized Laplace-Beltrami operator is highlighted. Red sites carry constant weights $ w^{(7,3)} $, whereas the central node (blue) is weighted by $ -7 w^{(7,3)}$.}
		\label{fig:tilings}
	\end{figure}

	Motivated by the goal to provide a new example of holographic duality, as well as possible realizations in tabletop experiments, in this Letter we report novel insights in this direction for discretized systems. A prime candidate is a scalar field defined on discretizations of AdS space via regular hyperbolic tilings (see Fig.~\ref{fig:tilings}) \cite{Coxeter,Magnus1974}, which have been recently investigated using methods from lattice gauge theory in \cite{PhysRevD.103.094507,Asaduzzaman,Asaduzzaman:2021bcw,Brower:2022atv}. These works consider discretization schemes for the scalar action, the Laplace operator, and lattice bulk propagators, finding good agreement of the scaling behavior of correlation functions with analytic continuum results.
	
	The physics of hyperbolic tilings has recently been studied in the context of condensed matter physics \cite{Bienias:2021lem}, circuit quantum electrodynamics \cite{Kollar2019,Kollar,Boettcher:2019xwl}, and topolectric circuits \cite{PhysRevLett.125.053901,Lenggenhager2021,Chen:2022}. These works focus on the spectrum of tight-binding Hamiltonians on hyperbolic lattices and their realization based on coupled waveguide resonators \cite{Kollar,Boettcher:2019xwl,Bienias:2021lem} or classical  non-dissipative linear electric circuits (topolectric circuits). Time-resolved measurements of wave propagation in hyperbolic space have been achieved in such architectures \cite{Lenggenhager2021}.
	
	It remains an open question, though, how to establish a duality in the sense of a map between  bulk and boundary theories for hyperbolic tessellations. Steps in this direction were taken in \cite{Axenides:2013iwa,Axenides:2019lea} using modular discretizations and in \cite{Gesteau:2022hss} via tensor networks on hyperbolic buildings. In this work we focus on hyperbolic tilings as a discretization scheme instead. Starting point is one of the key results of the \mbox{continuum} AdS/CFT correspondence, namely the relation between the mass $m$ of a scalar field in the bulk and the scaling dimension $\Delta$ of its dual operator on the boundary, $\Delta(\Delta-d)=m^2\ell^2$, with $\ell$ being the AdS radius and $d$ the boundary spacetime dimension \cite{Gubser:1998bc}. This is determined by the asymptotic boundary behavior of the solutions of the Klein-Gordon equation in AdS space.
	Conformally transforming AdS space into flat space, the scalar field experiences a shift of its mass-squared to $m^2 \ell^2 + d^2/4$.
	Thus, there is a stable potential minimum for the field if $m^2 \ell^2 > - d^2/4$, i.e. even for small negative mass-squared.
	This is the \textit{Breitenlohner-Freedman} (BF) bound \cite{BF1,BF2} \footnote{For a concise review of the BF bound, see \cite{Ammon:2015wua}, page 195.}.
	
	In this Letter, we determine how the BF bound is realized in discrete holographic setups and how it manifests itself on finite-sized architectures accessible through simulation and experiment.
	We establish the BF bound for hyperbolic tessellations by first analyzing the properties of the continuum  analytical solutions in the presence of a finite cutoff. This cutoff can be chosen arbitrarily and is required because only finite tilings, which do not cover the entirety of hyperbolic space, can be experimentally and numerically realized. In particular, this cutoff is independent of the Schläfli parameter $\{p,q\}$ characterizing a regular hyperbolic tiling with $q$ regular $p$-gons meeting at each vertex. Defining a scalar field on the vertices, we numerically solve the associated equations of motion on several tilings, finding excellent agreement with results from continuum holography. We find that the stability bound of a scalar field defined on large enough $\{p,q\}$ hyperbolic tilings coincides with the continuum BF bound, independently of $p$ and $q$. Our analysis extends previous investigations \cite{Kollar2019,Boettcher:2019xwl,Lenggenhager2021} of the eigenvalue problem of the discrete Laplacian on these tilings. In particular, we use insights from holography, such as the presence of non-normalizable modes, to provide solutions for masses-squared above the BF bound, thus beyond the standard spectrum of the Laplacian. Moreover, we propose a novel electric circuit, in the spirit of topolectric circuits \cite{RonnyTopolectric2018,ZhaoTopolectric2018}, to access these new mass-squared values in experiment.

	\paragraph{Equations of Motion on EAdS$_2$:}
	\label{sec:analytics}
	
	In order to investigate the physics of the BF bound for hyperbolic tilings, we consider one of the simplest continuum systems admitting a holographic duality, a free massive scalar field Euclidean AdS$_2$ (EAdS$_2$), with the induced metric
	\begin{equation}\label{eq:EAdS2}
		ds^2 = g_{\mu\nu} dx^\mu dx^\nu = \frac{\ell^2}{\cos^2(\theta)}\left(d\theta^2+\sin^2(\theta)d\phi^2\right)\,.
	\end{equation}
	Here, $\ell$ is the curvature radius of EAdS$_2$, and $\theta \in [0,\frac{\pi}{2})$, $\phi \in [0,2\pi)$. The asymptotic boundary of EAdS$_2$ is at $\theta = \frac{\pi}{2}$, corresponding to an infinite geodesic distance from the origin $ \theta=0 $. The scalar field action
	\begin{equation}\label{eq:action}
		S = \frac{1}{2} \int d^2 x \sqrt{g} \left(\partial^{\mu}\Phi\partial_{\mu}\Phi + m^2 \Phi^2\right)
	\end{equation}
	yields as equation of motion the Klein-Gordon equation \footnote{We denote \eqref{eq:Helmholtz1} as Klein-Gordon equation, although it does not contain a time derivative. We consider it to be fully defined by the metric \eqref{eq:EAdS2} and the Laplace-Beltrami operator.}
	\begin{eqnarray}\label{eq:Helmholtz1}
		0 &=& \frac{1}{\sqrt{g}}\partial_\mu \left(\sqrt{g}g^{\mu\nu} \partial_\nu \Phi\right) - m^2 \Phi\,\equiv (\square-m^2)\Phi\\
		&=& \frac{1}{\ell^2}\cos \theta \cot \theta \frac{\partial}{\partial\theta}\left(\sin \theta\, \frac{\partial \Phi}{\partial\theta}\right) -m^2 \Phi(\theta)\,,\label{eq:Helmholtz2}
	\end{eqnarray}
	with $\square$ the Laplace-Beltrami operator on EAdS$_2$. The second equality in \eqref{eq:Helmholtz2} holds for a purely $\theta$-dependent field configuration $\Phi(\theta)$ with no angular dependence.  Eq.~\eqref{eq:Helmholtz2} admits analytic solutions in terms of hypergeometric functions (S.2) in the Supplementary Material~\footnote{See Supplemental Material [URL inserted by the publisher] for more details.}, parametrized by two integration constants, which can be related by the regularity boundary condition $\Phi'(0)=0$.  Asymptotically near the boundary at $\theta=\pi/2$, the two fundamental solutions behave as
	\begin{equation}\label{eq:asymptoticbehavior}
		\Phi(\theta) \simeq A(\cos \theta)^{1-\Delta} + B (\cos \theta)^\Delta\,,
	\end{equation}
	where $\Delta = \frac{1}{2}+\sqrt{\frac{1}{4} + m^2\ell^2}$ is the scaling dimension of the boundary operator $O$ holographically dual to $\Phi$. In AdS/CFT, the terms in \eqref{eq:asymptoticbehavior} are denoted as \textit{non-normalizable} and \textit{normalizable} modes, respectively. Imposing suitable boundary conditions at $\theta=\frac{\pi}{2}$ \cite{Klebanov:1999tb}, the coefficient $B$ of the normalizable mode is identified with the vacuum expectation value of the dual operator $O$, while the coefficient $A$ of the non-normalizable mode determines its source $J$. In spatial dimension $ d=1 $ and for masses of the scalar field 
	\begin{align}
		m^2\ell^2 < -\frac{1}{4}\,,
		\label{eq:BFbound}
	\end{align} 
	the scaling dimension $\Delta$ of the dual operator becomes complex. In the CFT, this indicates a breakdown of unitarity. In the AdS bulk, it implies that the energy of the scalar field, 
	\begin{equation}
		E=\int d\theta \cos(\theta)^{2\Delta}\left((\partial_\theta\Tilde{\Phi})^2+\Delta^2\Tilde{\Phi}^2\right)\,,
		\label{eq:Energy}
	\end{equation}
	ceases to be a real and positive quantity \cite{Note3}. Here, we have used the ansatz $\Phi(\theta)=(\cos\theta)^{\Delta}\Tilde{\Phi}(\theta)$ \cite{BF1,BF2,MEZINCESCU1985406}. After quantization, this denotes an instability of the system towards a new, true ground state \cite{Kaplan:2009kr}. The reality condition on $\Delta$ implies $ m^2\ell^2 \geq -\frac{1}{4} $. This is the Breitenlohner-Freedman stability bound \cite{BF1,BF2} and is a key element of AdS/CFT.
	
	\begin{figure}[t]
		\centering
		\includegraphics[width=\columnwidth]{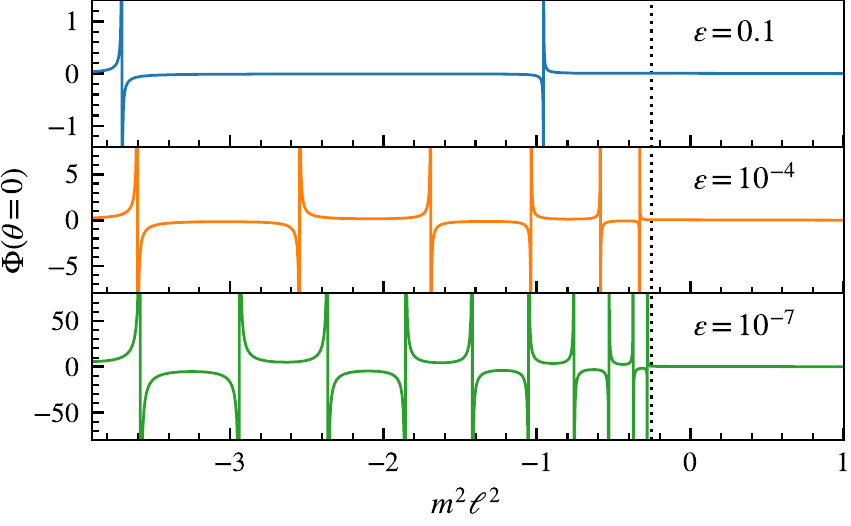}
		\caption{Field amplitude at the origin for different cutoff values $\varepsilon$. We observe \textit{Umklapp~points} appearing at any finite cutoff, the rightmost of which can be used as an indicator of unstable solutions. For smaller cutoffs, the Umklapp~points become denser and converge towards the continuum BF bound (dotted line).}
		\label{fig:umklapp}
	\end{figure}

	We now analyze the implications of a finite cutoff for the analytical solutions. Regularity at the origin is imposed through von Neumann boundary conditions at $\theta=0$. Introducing a finite radial cutoff $\varepsilon\ll 1$, such that Dirichlet boundary conditions are imposed at $\theta_c=\frac{\pi}{2}-\varepsilon$ via $\Phi(\theta_c)=1$, leads to a rescaling of the solutions. We solve Eq.~\eqref{eq:Helmholtz2} subject to these boundary conditions and find that, above the BF bound, solutions have no zeroes in the regime $\theta\in[0,\frac{\pi}{2})$. Below the BF bound, however, solutions develop an infinite set of zeroes \cite{Note3}.
	
\begin{figure}[t]
		\centering
		\includegraphics[width=\columnwidth]{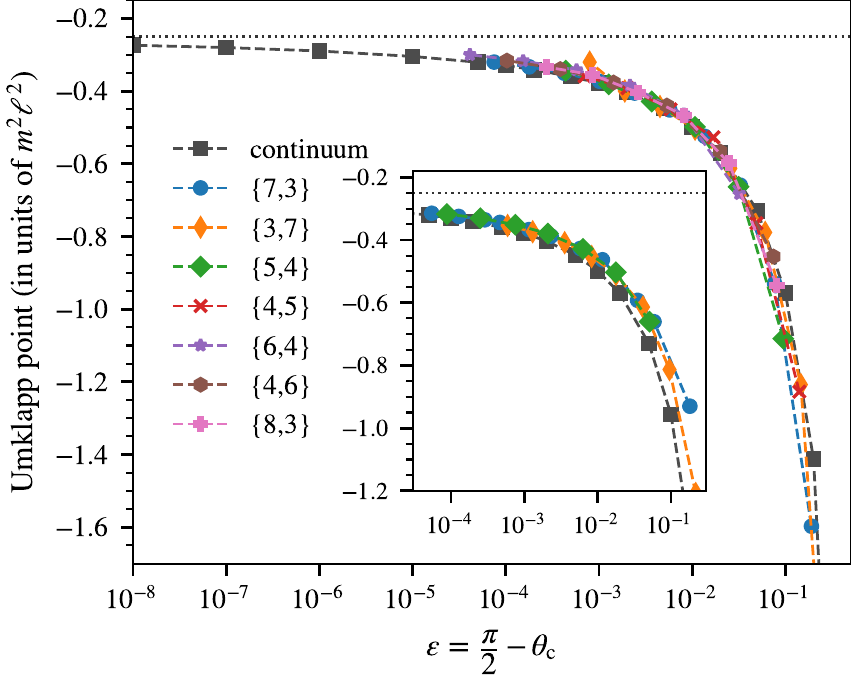}
		\caption{First Umklapp~point for various hyperbolic tilings and radial cutoffs $\theta_c$. For small $\varepsilon$, the curves tend towards the continuum bound $ m^2\ell^2=-1/4 $, indicated by the dotted horizontal line.
			Inset: corresponding results from our hyperbolic electric circuit simulations.}
		\label{fig:bound}
	\end{figure}

	At specific values of the mass-squared $m^2\ell^2$ and cutoff $\varepsilon$, we observe a singular behavior of the solutions, characterized by discontinuous jumps of the field amplitude, as presented in Fig.~\ref{fig:umklapp}. These only appear below the BF bound and are a result of the cutoff coinciding with a zero of the solution associated to the given value of $m^2\ell^2$, thus making the rescaling factor diverge. We denote these pairs $(m^2\ell^2,\varepsilon)$ as \textit{Umklapp~points}. The position of the first Umklapp~point below $m^2\ell^2=-\frac{1}{4}$, corresponding to the zero which is furthest into the bulk, can be used as an indicator for the unstable regime. More precisely, when the cutoff is removed, the value of the mass-squared for which this zero first appears corresponds to the BF stability bound. The analytical derivation of the solutions to the KG equation \eqref{eq:Helmholtz2} at a finite cutoff $\varepsilon$ provided in \cite{Note3} allows for an exact tracking of this first Umklapp~point for different cutoffs. This provides a reference behavior, shown in black in Fig.~\ref{fig:bound}, to which we compare our numerical findings.
	
	\paragraph{Regular hyperbolic tilings of $\mathds{D}^2$:}
	\label{sec:tilings}
	
	EAdS$_2$ is isomorphic to the Poincar\'e disk model of hyperbolic space $\mathds{D}^2$, which can be naturally discretized by \textit{regular hyperbolic tilings} \cite{Coxeter,Magnus1974}. These preserve a large subgroup, known as a \textit{Fuchsian group of the first kind}, of the isometry group PSL$(2,\mathds{R})$ of EAdS$_2$ \cite{katok1992fuchsian,Boettcher:2021njg}, making them promising candidates for setting up a discrete holographic duality. Hyperbolic tilings are characterized by their Schläfli symbol $\{p,q\}$, with $(p-2)(q-2)>4$, denoting a tiling with $q$ regular $p$-gons meeting at each vertex. The $\{7,3\}$ hyperbolic tiling and its dual $\{3,7\}$ tiling are shown in Fig.~\ref{fig:tilings} as an example. Since hyperbolic space introduces a length scale through its radius of curvature $\ell$, the edge lengths of hyperbolic polygons are fixed quantities, depending only on the Schläfli parameters $p$ and $q$ \cite{coxeter_1997}. Their geodesic length $\theta^{(p,q)}$ in units of $\ell$ can be computed via the Poincaré metric \eqref{eq:EAdS2} and can be interpreted as a fixed lattice spacing that cannot be tuned. We compute $\theta^{(p,q)}$ for several $p$ and $q$ in \cite{Note3}. In general, this makes a continuum limit of $\{p,q\}$ tilings in the usual way impossible. Nevertheless, we provide evidence that regular hyperbolic tilings indeed preserve some properties of the continuum scalar field theory, indicating that they are a good approximation of continuum EAdS$_2$.
	
	While the entire EAdS$_2$ space can be filled with an infinite $\{p,q\}$ tiling, numerical simulations and experimental setups can only be finite-sized. The truncation of the tiling to a finite number of layers is equivalent to the introduction of a finite cutoff as mentioned earlier. Given the jagged structure of the tiling's boundary at any finite layer, an effective uniform radial cutoff needs to be drawn. This allows a direct comparison of the Umklapp points observed in numerical simulations on the truncated tilings with the analytical solutions derived in \cite{Note3}.

	\paragraph{Numerical Methods:}
	
	The central ingredient for our numerical analysis of the Klein-Gordon equation on $\{p,q\}$ hyperbolic tilings is a suitable discretization, denoted by~$\tilde \square$, of the Laplace-Beltrami operator. Its action on a scalar function $\Phi(\theta,\phi)$, represented on the tiling by discrete values $\Phi_j = \Phi(\theta_j)$, can be written as
	\begin{align}
		\bigl( \tilde \square\, \Phi \bigr)_j = \sum_{k|j} w_{jk} \ell^{-2} (\Phi_k-\Phi_j),
		\label{LaplaceDiscretized}
	\end{align}
	where $k|j$ denotes the summation over the $q$ neighboring sites $k$ of site $j$. In order to determine the weight factors $w_{jk}$, we implement the following method devised from the established approximation of lattice operators by finite difference quotients. Given that all cells in a hyperbolic tiling are isometric, it is clear that all weights in the stencil have to be equal, \ie $w_{j,k}\equiv w$ (cf. right panel of Fig.~\ref{fig:tilings}), which allows us to write the lattice operator formally as a matrix
	\begin{align}
		\tilde \square-m^2  \equiv w \ell^{-2} (A-G) + M,
		\label{eq:LaplaceMatrix}
	\end{align}
	acting on a vector of function values $\Phi=(\Phi_0, \Phi_1, \ldots)$. Here, $A$ and $G$ denote the adjacency and degree matrix of the tiling graph and $M=\mathrm{diag}(-m^2)$. In order to calculate the weight $w$, recall the approximation of a 1D second derivative by finite differences $\tilde \partial_x^2\chi(x) = w (\chi (x-h)-2 \chi (x)+\chi (x+h))$. First, we determine a test function $\chi (x)$ such that $\partial_x^2 \chi (x)=1$, in this case $\chi (x)=\frac12 x^2$. Applying the discretized derivative to this function yields  $\tilde\partial_x^2 \chi (x)=h^2 w$, hence $w=h^{-2}$. Note that unlike this case of a Cartesian hypercubic lattice, the hyperbolic lattice spacing $h$ is fixed, as discussed above.
	
	We now apply this procedure to the stencil on the hyperbolic lattice. First, we determine a radially symmetric test function $\chi(\theta)$ such that
	\begin{equation}
		\square\chi(\theta)= \frac{1}{\ell^2}\Bigl( \cos^2\theta \cot\theta\,\partial_\theta+\cos^2\theta \,\partial_\theta^2  \Bigr)\chi(\theta)\;=\; 1\,.
	\end{equation}
	A possible solution is
	\begin{equation}
		\chi(\theta)=\ell^2\, \ln\Bigl( 1+\frac{1}{\cos\theta} \Bigr)\,.
	\end{equation}
	Applying $ \tilde\square $ to this function on the central site of the tiling (cf. right panel of Fig.~\ref{fig:tilings}) yields
	\begin{equation}
		\tilde \square \chi(0) = p\,w\Bigl(\chi(\theta^{(p,q)})-\chi(0)\Bigr) = 1.
		\label{eq:weightformula}
	\end{equation}
	Solving for $w$, the weight factors can be obtained for every $\{p,q\}$ and are listed in \cite{Note3}.

	\paragraph{Numerical Results:}
	
	Given the lattice Laplacian, the continuum Klein-Gordon equation on a constant time slice can be expressed on the finite hyperbolic tiling as
	\begin{align}
		\begin{split}
			\tilde{\square}\,\Phi-m^2\Phi &= 0 \quad \text{for}\quad  \theta<\theta_c,\\
			\Phi(\theta)&=\Phi_c \quad \text{for}\quad \theta=\theta_c\,
		\end{split}
		\label{eq:boundaryvalueproblem}
	\end{align}
	where the boundary condition is implemented by assigning constant values to sites outside the radial cutoff $\theta_c$. 
	
	Solving the discretized boundary value problem \eqref{eq:boundaryvalueproblem} requires iterative matrix methods~\cite{asmar2016,morton2005} already for medium system sizes. It has to be taken into account that for negative $m^2$ exceeding a certain threshold, most standard solvers tend to be unstable due to the lattice operator $ \tilde{\square}-m^2 $ becoming indefinite in this parameter regime \cite{ernst2012}. A class of algorithms which can handle indefinite, sparse linear systems are so-called Krylow subspace methods~\cite{saad2003}. In particular, we use both the GMRES (generalized minimum residual)~\cite{saad1986} and BiCGSTAB (biconjugate gradient stabilized)~\cite{van1992} methods to solve Eq.~\eqref{eq:boundaryvalueproblem} and extract the positions of the Umklapp points. Results from both algorithms are fully compatible and presented in Fig.~\ref{fig:bound} for various hyperbolic tilings~\cite{schrauth2022} and values of the cutoff. We find that all curves nicely converge towards $m^2\ell^2=-1/4$ for $\varepsilon\to 0$, thus yielding the correct infinite volume limit and marking the main result of this Letter.

	We expect that the universal behavior for all $p$ and $q$ displayed in Fig.~\ref{fig:bound} originates from a group-theoretic argument as follows. For Fuchsian groups of the first kind, which describe the isometries of $\{p,q\}$ tilings, it is known that the boundary limit set is the circle $S^1$ \cite{katok1992fuchsian,Gesteau:2022hss}. For infinite tilings, this implies conformal invariance of the boundary theory. The BF bound is the mass-squared threshold at which the scaling dimension $\Delta$ of the CFT operator dual to the bulk scalar field becomes complex, as can be seen from the definition of $\Delta$ below \eqref{eq:asymptoticbehavior}. Thus, the asymptotic value of the first Umklapp~point must be the same for all $p$ and $q$ as $\varepsilon\to0$. In addition, our numerical results of Fig.~\ref{fig:bound} indicate that even for finite cutoff, where the Fuchsian symmetry is broken, the universality of the $\varepsilon\to0$ behavior is preserved for all $p$ and $q$. 
	
	\paragraph{Hyperbolic Electric Circuits:}\label{sec:topolectric}
	
	\begin{figure}[t]
		\includegraphics[width=0.90\columnwidth]{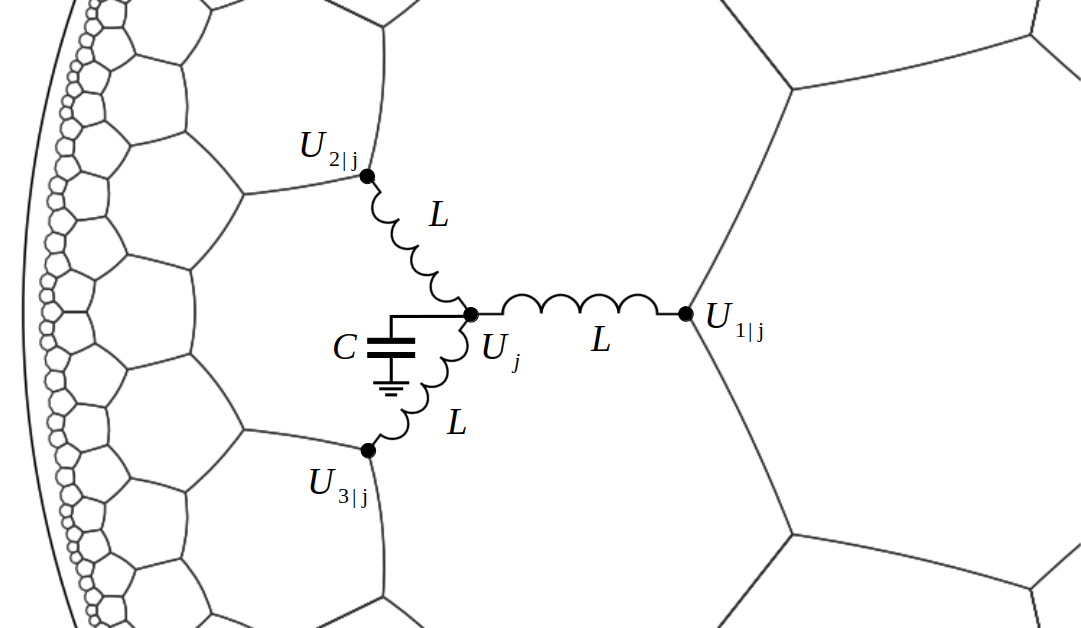}
		\caption{Section of the hyperbolic electric circuit. The structure is repeated at every vertex in the lattice.} 
		\label{fig:topolectric}
	\end{figure}

	We further propose an experimental realization of the BF bound in a suitable electric circuit. We are motivated by topolectric circuits \cite{PhysRevLett.125.053901}, which are a platform based on circuits of capacitors and inductors which are engineered to realize a plethora of models exhibiting topological states of matter~\cite{RonnyTopolectric2018,ZhaoTopolectric2018,dong2021topoelectric}. Specifically, let us consider a circuit on a hyperbolic tiling as shown in Fig.~\ref{fig:topolectric}. On the vertices of the tiling we attach grounded capacitors $C$ and connect them via identical inductors $L$ along the polygon edges. Note that our construction differs from that in \cite{Lenggenhager2021} in that we exchange capacitors and inductors. This is necessary because only then the site voltage $U_j$ represents the scalar field $\phi_j$. In this network, the voltage $U_j$ at site $j$ is related to the capacitor current by $I_j=C\dot U_j$ while the voltage differences between neighboring sites $k$ of $j$ are related to the induced current by $(U_{k}-U_{j})=L \dot I_{jk}$. According to Kirchhoff's laws, the time-evolution of the voltage at site $j$ is given by
	\begin{align}
		\ddot U_j = \frac{1}{LC}\sum_{k|j} (U_{k}-U_{j}),
	\end{align}
	where in our case, the weights are constant $ w_{jk} = w^{(p,q)} $ as discussed above. The oscillatory eigenmodes $U_j(t) = u_j e^{i \omega t}$ are determined by the system of equations 
	\begin{align}
		-\omega^2 U_j =\frac{1}{LC}\sum_{k|j} (U_{k}-U_{j}) = \frac{w^{(p,q)}}{L C}\tilde \square U_j\,.
		\label{eq:CircuitEquationSystem}
	\end{align}
	By identifying 
	\begin{align}
		-m^2\ell^2=\omega^2 \frac{LC}{w^{(p,q)}},
		\label{eq:DispersionRelation}
	\end{align}
	the electric circuit provides a realization of the discretized Klein-Gordon equation for $m^2\ell^2<0$ \footnote{Note that the time $t$ is an auxiliary parameter introduced to parametrize the oscillatory eigenmodes of the circuit and is not related to the total energy \eqref{eq:Energy} of the gravitational system.}.

	The first Umklapp~point corresponds to the lowest eigenfrequency of the circuit, which represents the finite gap in the negative-definite eigenspectrum of the hyperbolic Laplacian \cite{Boettcher:2019xwl}. Since the circuits described above contain only passive elements, they can only realize the regime of negative mass-squared. This is however precisely the regime where according to \eqref{eq:Energy}, solutions to \eqref{eq:Helmholtz2} are unstable within the AdS/CFT correspondence. For electric circuits to access the regime of $m^2$ above the BF bound, thus realizing non-normalizable solutions \eqref{eq:asymptoticbehavior} essential for holography, the implementation of active electrical elements is required. Such elements were introduced in \cite{RonnyActiveTopo} in the context of topolectric circuits. We propose to use negative impedance converters to achieve negative values of $L$ or $C$ on the r.h.s.~of \eqref{eq:DispersionRelation}. 
	
	In order to systematically locate the eigenmodes of the passive circuit, we apply a driving alternating current at the central node and integrate the system of Eqs.~\eqref{eq:CircuitEquationSystem} over time using an explicit fourth-order Runge-Kutta method. Details of our numerical analysis are presented in \cite{Note3}. Once the fundamental mode of the system is found, the corresponding negative mass-squared can be extracted according to Eq.~\eqref{eq:DispersionRelation}. These resonances (eigenfrequencies of the circuit) are a physical manifestation of the Umklapp~points introduced above. Performing this analysis for several different hyperbolic tilings~\cite{schrauth2022} and finite cutoff radii, we are able to locate the positions of the lowest eigenfrequency. Similarly to our analysis of the Umklapp~points, we are able to find the instability threshold on the tiling by tracking the position of the first resonance frequency of the circuit as the cutoff is removed. Again, we find an excellent agreement with the continuum prediction, as shown in the inset of Fig.~\ref{fig:bound}. Our analysis thus shows how the BF bound can be experimentally realized on hyperbolic electric circuits.

	\paragraph{Conclusions:}
	
	For the first time, we have identified the implications of the Breitenlohner-Freedman bound for discrete regular tilings of hyperbolic space. Notably, we find universal behavior of the instabilities for all $\{p,q\}$ discretizations, even for finite cutoff. In particular, we find excellent agreement between the positions of the Umklapp~points as obtained via numerical simulations of the scalar field on several different $\{p,q\}$ tilings with the analytical solutions of the Klein-Gordon equation on EAdS$_2$. Moreover, for a specific hyperbolic electric circuit we show how the resonance frequencies are a manifestation of the Umklapp~points. Simulations of the circuit dynamics also show excellent agreement with the analytical data by yielding the same dependence of the resonances on the cutoff size. Both these results confirm the universal behavior. Furthermore, we suggest how to adapt the electrical circuits in order to realize mass-squared values above the BF bound. Such circuit realizations will make regular hyperbolic tilings excellent candidates for bringing aspects of AdS/CFT to the laboratory.
	
	The EAdS$_2$ manifold considered here describes a constant time slice of the larger AdS$_{2+1}$ spacetime. It would be interesting to generalize our analysis to a Lorentzian setting involving time, for instance by adding a temporal leg to the vertices of the tilings and equipping them with radius-dependent weights (see also \cite{Brower:2022atv} for a first attempt in this direction). In practice, this can be implemented by locally modifying $L$ and $C$ on the hyperbolic electric circuit. We leave this for future work.

\newpage

\begin{acknowledgments}
	\textbf{Acknowledgements:} We are grateful to R.~Thomale, R.~N.~Das and G.~Di~Giulio for fruitful discussions. Moreover, we thank F.~Goth for helpful discussions regarding the numerical aspects of this article. P.B., J.E., R.M. and M.S.~acknowledge support by the Deutsche Forschungsgemeinschaft (DFG, German Research Foundation) under Germany's Excellence Strategy through the W\"urzburg‐Dresden Cluster of Excellence on Complexity and Topology in Quantum Matter ‐ ct.qmat (EXC 2147, project‐id 390858490). J.E. and R.M.~furthermore acknowledge financial support through the  Deutsche  Forschungsgemein\-schaft  (DFG,  German  Research  Foundation),  project-id   258499086   -   SFB   1170   ’ToCoTronics'.
\end{acknowledgments}

\onecolumngrid
\renewcommand{\theequation}{S.\arabic{equation}}
\vspace*{20mm}

\section{Supplementary Material}

\subsection{Analytic Solutions on EAdS$_2$}\label{sec:AppA}

We provide the full analytical derivation of the solutions to the Klein-Gordon equation in EAdS$_2$. In the coordinates given in Eq.~(1) of the main text, the Klein-Gordon equation in Eq.~(4) can be written as
\begin{equation}
	\cos(\theta)^2\partial^2_{\theta}\Phi(\theta)+\cos(\theta)^2\cot(\theta)\partial_{\theta}\Phi(\theta)-m^2\ell^2\Phi(\theta)=0\,.
	\label{KGEquation}
\end{equation}
This differential equation has two fundamental solutions in terms of hypergeometric functions \cite{abramowitz1965handbook}, whose linear combination gives the general solution
\begin{multline}
	\Phi(\theta)=(-1)^{\frac{1}{4}-\frac{1}{4}\sqrt{1+4m^2\ell^2}}\cos(\theta)^{\frac{1}{2}-\frac{1}{2}\sqrt{1+4m^2\ell^2}}\\
	\times\Big(\tilde{A}\cdot _{\,2}F_1[a_1,b_1,c_1;\cos(\theta)^2]
	+\tilde{B}\left(i\,\cos(\theta)\right)^{\sqrt{1+4m^2\ell^2}}\cdot _{\,2}F_1[a_2,b_2,c_2;\cos(\theta)^2]\Big)\,,
	\label{eq:FullSolution}
\end{multline}
with the coefficients
\begin{equation}
	\begin{split}
		a_1&=\frac{1}{4}\left(1-\sqrt{1+4m^2\ell^2}\right)\,,\\
		b_1&=\frac{1}{4}\left(3-\sqrt{1+4m^2\ell^2}\right)\,,\\
		c_1&=\frac{1}{2}\left(2-\sqrt{1+4m^2\ell^2}\right)\,,\\
		a_2&=\frac{1}{4}\left(1+\sqrt{1+4m^2\ell^2}\right)\,,\\
		b_2&=\frac{1}{4}\left(3+\sqrt{1+4m^2\ell^2}\right)\,,\\
		c_2&=\frac{1}{2}\left(2+\sqrt{1+4m^2\ell^2}\right)\,.
	\end{split}
	\label{eq:HypergeometricCoefficients}
\end{equation}
$\tilde{A}$ and $\Tilde{B}$ are integration constants. Neglecting for the moment constant prefactors we can already extract the asymptotic behavior of the solutions \eqref{eq:FullSolution} towards the boundary $\theta\rightarrow\frac{\pi}{2},\,\cos(\theta)^2\rightarrow 0$. For this, recall that the hypergeometric functions are regular around the origin $z=0$, with their series expansion given by \cite{abramowitz1965handbook}
\begin{equation}
	_{\,2}F_1[a,b,c;z]=1+\frac{a b z}{c}+\frac{a (a+1) b (b+1) z^2}{2 c (c+1)}+\mathcal{O}\left(z^3\right)\,.
\end{equation}
Thus, the solutions \eqref{eq:FullSolution} have the asymptotic boundary behavior
\begin{equation}
	\Phi(\theta)\overset{\theta\rightarrow\frac{\pi}{2}}{\approx} A \cos(\theta)^{1-\Delta}\left(1+\mathcal{O}(\theta^2)\right)+B \cos(\theta)^{\Delta}\left(1+\mathcal{O}(\theta^2)\right)
\end{equation}
with $\Delta=\frac{1}{2}+\frac{1}{2}\sqrt{1+4m^2\ell^2}$ and where the coefficients $A,\,B$ include possible constant prefactors.\\
The solutions \eqref{eq:FullSolution} diverge at the origin $\theta\rightarrow 0$ due to individual logarithmic divergences of the hypergeometric functions at that singular point. Thus, to guarantee regularity of the solutions at the origin, i.e. $\partial_{\theta}\Phi(\theta)|_{\theta=0}=0$, we need to specify a relation between the integration constants $\Tilde{A}$ and $\Tilde{B}$ such that the logarithmic divergences cancel each other. A straightforward computation at $\theta=0$ yields 
\begin{equation}
	\tilde{A}=-\frac{(i)^{\sqrt{1+4m^2\ell^2}}\,_{\,2}F_1[a_2,b_2,c_2;1]}{ _{\,2}F_1[a_1,b_1,c_1;1]}\Tilde{B}\,.
	\label{eq:FirstRegularityCond}
\end{equation}
This relation can be further simplified by exploiting the properties of hypergeometric functions with unity argument \cite{abramowitz1965handbook}
\begin{equation}
	\begin{split}
		_{\,2}F_1[a,b,c;1]=\begin{cases}
			\frac{\Gamma(c)\Gamma(c-a-b)}{\Gamma(c)\Gamma(c-b)}\quad ; \mathfrak{Re}(c-a-b)>0\,,\\
			\Lim{z\rightarrow 1^-}\frac{ _{\,2}F_1[a,b,c;z]}{-\ln(1-z)}=\frac{\Gamma(a+b)}{\Gamma(a)\Gamma(b)}\quad ; c=a+b\,,
		\end{cases}
	\end{split}
\end{equation}
where $\Gamma$ denotes the Euler-Gamma function. The coefficients \eqref{eq:HypergeometricCoefficients} obey $c_i=a_i+b_i$ for $i=1,2$ and we can thus simplify \eqref{eq:FirstRegularityCond} to find
\begin{equation}
	\tilde{A}=-(i)^{\sqrt{1+4m^2\ell^2}}\frac{\Gamma(a_2+b_2)\Gamma(a_1)\Gamma(b_1)}{\Gamma(a_1+b_1)\Gamma(a_2)\Gamma(b_2)}\tilde{B}\,.
\end{equation}
Finally, we wish to impose Dirichlet boundary conditions at a finite radial cutoff $\theta=\theta_c=\frac{\pi}{2}-\varepsilon$, such that the field value equals unity, i.e. $\Phi(\theta_c)=1$. This fixes the remaining integration constant to be
\begin{equation}
	\tilde{B}=\frac{(-1)^{b_2} \Gamma \left(c_1\right) \Gamma \left(c_2\right) 2^{\sqrt{4 \ell^2 m^2+1}} \cos ^{\Delta -1}(\theta_c)}{\sqrt{\pi } \left(\Gamma \left(4 a_1\right) \Gamma \left(c_1\right) (8 i)^{\sqrt{4 \ell^2 m^2+1}} \, _2\tilde{F}_1\left[a_1,b_1,c_1;\cos ^2(\theta_c)\right]-\Gamma \left(4 a_2\right) \Gamma \left(c_1\right) (i \cos (\theta_c))^{\sqrt{4 \ell^2 m^2+1}} \, _2\tilde{F}_1\left[a_2,b_2,c_2;\cos ^2(\theta_c)\right]\right)}\,,
\end{equation}
with the regularized hypergeometric function $_2\tilde{F}_1[a,b,c;z]=\frac{_2F_1[a,b,c;z]}{\Gamma(c)}$. The solutions are thus fully determined as functions of the mass $m^2$ and the radial cutoff $\theta_c=\frac{\pi}{2}-\varepsilon$.

\subsection{Positivity of the Energy Functional}\label{sec:AppB}

\begin{figure}[t]
	\vspace*{5mm}
	\includegraphics[width=0.9\columnwidth]{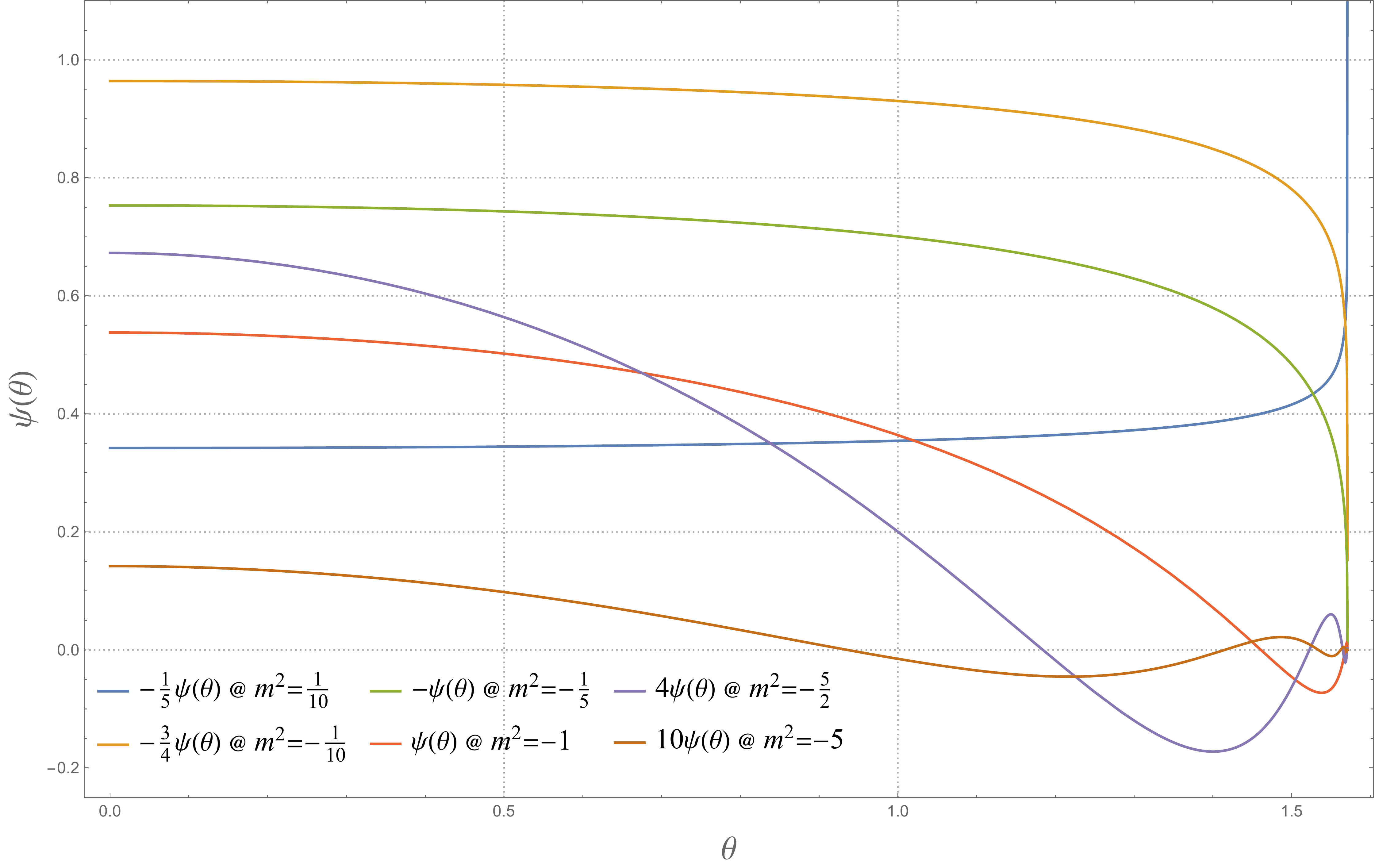}
	\caption{Comparison of solutions to the Klein-Gordon equation for values of the mass-squared above and below the BF bound $m^2 \ell^2 = - 1/4$. The solutions have been rescaled for clarity. The solutions above the BF bound for $m^2 \ell^2=-1/10$ and $m^2 \ell^2 =-1/5$ do not exhibit any zeroes, while the solutions below the BF bound $m^2<-1/4$ do. For positive values of the mass-squared, the solutions contain non-normalizable modes that diverge at the boundary $\theta\to\pi/2$ (blue curve).} 
	\label{fig:comparisonplot}
\end{figure}

In this section we show how solutions to the Klein-Gordon equation for masses below the BF bound are unstable. This instability is characterized by what was denoted as positivity of the energy in the original work of Breitenlohner and Freedman \cite{BF1,BF2}. In their context, positivity requires the energy to be a a positive real number. If the energy is negative or develops a non-vanishing imaginary part, the system is considered to be unstable. The arguments in this section have been provided partly in the original papers by Breitenlohner and Freedman \cite{BF1,BF2}, as well as in an extended analysis given in \cite{MEZINCESCU1985406}. These works considered the Lorentzian version of AdS$_d$ spacetime in higher dimensions, but the logical arguments are insensitive to a Wick rotation and the choice of $d=2$.\\
The scalar action 
\begin{equation}
	S=\frac{1}{2}\int d^2x\sqrt{g}\left(\partial^{\mu}\Phi\partial_{\mu}\Phi+m^2\Phi^2\right)\,,
\end{equation}
with $\mu,\nu=\{\theta,\phi\}$ in the coordinates of Eq.~(1) of the main text. We can associate a corresponding stress-energy tensor $T_{\mu\nu}$ through the variation
\begin{equation}
	T_{\mu\nu}\equiv \frac{2}{\sqrt{g}}\frac{\delta S}{\delta g^{\mu\nu}}\,.
\end{equation}
For our scalar field, we obtain explicitly
\begin{equation}
	T_{\mu\nu}=\partial_{\mu}\Phi\partial_{\nu}\Phi+\frac{1}{2}g_{\mu\nu}\left(g^{\rho\sigma}\partial_{\rho}\Phi\partial_{\sigma}\Phi+m^2\Phi^2\right)\,.
	\label{eq:StressEnergyTensor}
\end{equation}
The energy functional of the full system decomposes into a contribution from gravity and a contribution from the scalar field. The positivity of the gravity contribution was proven in \cite{ABBOTT198276}. The contribution from the scalar field is given by the energy functional
\begin{equation}
	E(T_{\mu\nu})=\int_0^{\frac{\pi}{2}}d\theta\, T_{\phi\phi}\,,
\end{equation}
where the integrand denotes the \enquote{temporal} component of the stress energy tensor. In Euclidean signature, this corresponds to the angular coordinate $\phi$ present in Eq.~(1) of the main text. Inserting our stress-energy tensor \eqref{eq:StressEnergyTensor}, we obtain
\begin{equation}
	E(T_{\mu\nu})=	\int_0^{\frac{\pi}{2}}d\theta \left(\frac{3}{2}(\partial_{\phi}\Phi)^2+\frac{1}{2}(\partial_{\theta}\Phi)^2+\frac{1}{2}\frac{m^2\ell^2}{\cos(\theta)^2}\Phi^2\right)\,.
	\label{eq:Energyfunctional}
\end{equation}
This expression might seem manifestly positive, but the potential term can contribute negatively for negative mass-squared. Since we are interested radially symmetric solutions, we can neglect the first term. This is not a restriction, since this term by itself is indeed manifestly real and positive. We will thus not write it explicitly in the following. In order to re-write the radial derivative and the potential terms in manifestly real and positive form, we use the ansatz $\Phi(\theta)=\cos(\theta)^{\Delta}\Tilde{\Phi}$, where the exponent is \textit{a priori} not specified but we denoted it with a suggestive notation in view of the scaling dimension. Plugging in this ansatz into \eqref{eq:Energyfunctional} and integrating by parts, we find
\begin{equation}
	E(T_{\mu\nu})= \int_0^{\frac{\pi}{2}} d\theta \cos(\theta)^{2\Delta}\left((\partial_{\theta}\Tilde{\Phi})^2+\left(\Delta+\frac{\Delta(1-2\Delta)\sin(\theta)^2+\Delta^2+m^2\ell^2}{\cos(\theta)^2}\right)\Tilde{\Phi}^2\right)-\left[2\Delta\cos(\theta)^{2\Delta-1}\sin(\theta)\Tilde{\Phi}^2\right]_0^{\frac{\pi}{2}}\,.
\end{equation}
The vanishing of the boundary term for normalizable modes was proven in \cite{BF1,BF2} and for non-normalizable modes in \cite{MEZINCESCU1985406}. It thus remains to guarantee the reality and positivity of the integrand. By choosing the coefficient in our ansatz to obey $\Delta=\frac{1}{2}+\sqrt{\frac{1}{4}+m^2\ell^2}$, we can write
\begin{equation}
	E(T_{\mu\nu})= \int_0^{\frac{\pi}{2}} d\theta \cos(\theta)^{2\Delta}\left((\partial_{\theta}\Tilde{\Phi})^2+\Delta^2\Tilde{\Phi}^2\right)\,.
\end{equation}
This expression is manifestly real and indeed positive as long as $\Delta$ is real, which in turn requires $m^2\ell^2\geq-\frac{1}{4}$. This is precisely the BF bound. For visualization purposes, the behavior of the solutions above and below the BF bound is shown in Fig.~\ref{fig:comparisonplot}. We observe a drastic change of the solutions when the mass-squared goes below the stability threshold.

\subsection{Hyperbolic Electric Circuits}

\begin{table}[b]
	\centering
	\begin{tabularx}{0.4\columnwidth}{C{16mm}C{22mm}C{22mm}} 
		\hline
		\hline
		$\{p,q\}$ & $ w^{(p,q)} $ & \Vghost$ \theta^{(p,q)} $\Vghost\\
		\hline
		$\{3,7\}$ & 4.213387 & 0.5382155\\
		$\{4,5\}$ & 0.928414 & 0.9045569\\
		$\{4,6\}$ & 0.616576 & 1.0471976\\
		$\{5,4\}$ & 0.541067 & 1.0147126\\
		$\{6,4\}$ & 0.240449 & 1.2309594\\
		$\{7,3\}$ & 0.503616 & 0.9224420\\
		$\{8,3\}$ & 0.233732 & 1.1437177 \\
		\hline \hline
	\end{tabularx} 
	\caption{Weights and edge lengths for the tilings used in this paper.}
	\label{tab:weights}
\end{table}

Here we describe in detail our numerical simulations regarding a possible experimental realization of the Breitenlohner-Freedman bound, using an electric circuit network. The primary goal is to locate the fundamental oscillatory eigenmode of the circuit. The corresponding resonance frequency directly corresponds to the position of the stability bound via Equation~(15) in the main text. Specifically, we implement the setup shown in Fig.~4 in the main text numerically. The equations governing the dynamics of the currents along the links and voltages on the nodes are given by
\begin{align}
	\begin{split}
		L \dot I_{jk} &= w^{(p,q)} (U_{k}-U_{j})\\
		C\dot U_j &= I_j,
	\end{split}
	\label{eq:Kirchhoff}
\end{align}
where the dot denotes a temporal derivative. This represents a system of $2N$ coupled first-order differential equations, where $N$ is the number of nodes in the tiling and $ j,k $ are vertex indices. The system is to be solved for the time-dependent voltages $U_j(t)$, $j=1,\ldots,N$. In our simulations, we set $ L=C=1 $. The weights $ w^{(p,q)} $ are given in Table \ref{tab:weights} for a selection of $ (p,q) $ pairs. Note that in our actual technical realization, these weights are introduced as the relative strength of the inductances as compared to the capacitances.

In order to integrate Equations~\eqref{eq:Kirchhoff} over time, we use an explicit fourth-order Runge-Kutta (RK4) method~\cite{atkinson1989,press2007}. Radial Dirichlet boundary conditions are realized by first truncating the hyperbolic lattice at a fixed radius $\theta_c<\frac{\pi}{2}$. Then, only those nodes inside the cutoff radius are iterated according to the dynamic rules \eqref{eq:Kirchhoff}. Sites which are positioned outside of the cutoff are considered ghost nodes. They are grounded without an additional capacitor and hence effectively not updated during the simulation. Consequently, their voltages remain zero, \ie $ U_j(t) = 0 $ for $\theta(x_j)>\theta_c$. We prepare the system in a state where all voltages are set to zero initially and apply a driving alternating voltage at the central site of the tiling, $ U(x_0,t) \equiv U_0(t) \propto \sin(\omega_d t) $. Note that in order to resolve the temporal dynamics of the circuit accurately, the time increment of the integrator is chosen significantly smaller than the typical time scale of the system, \ie $ \Delta t \ll \omega^{-1} $. 

\begin{figure}[t]
	\includegraphics[width=0.55\columnwidth]{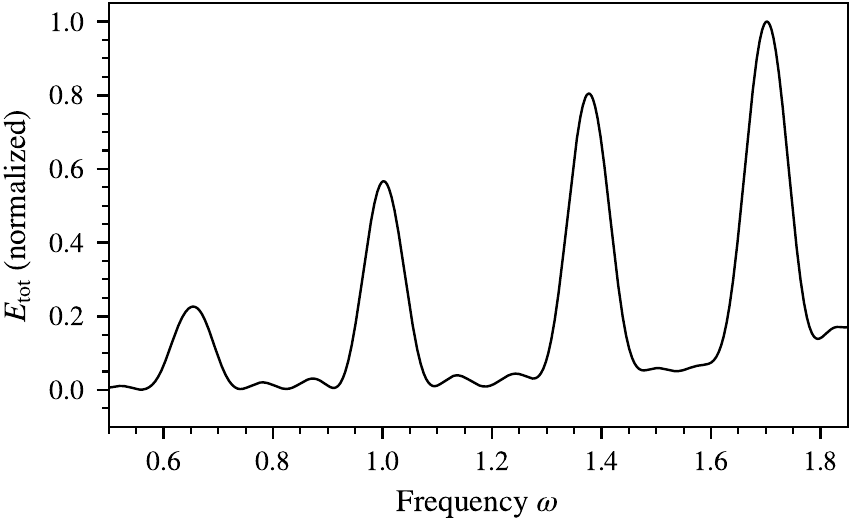}
	\caption{Normalized total energy $E_{\textrm{tot}}$ of the circuit as defined in \eqref{eq:total_energy}, as a function of the driving frequency, measured after the shut down of the drive.} 
	\label{fig:energy_spectrum}
\end{figure}

\begin{figure}[t]
	\vspace*{5mm}
	\includegraphics[width=0.46\columnwidth]{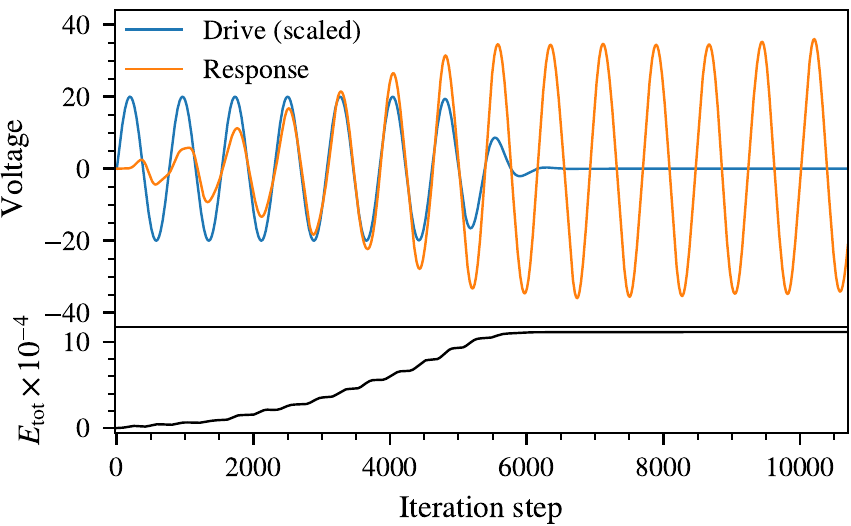}\hspace*{5mm}
	\includegraphics[width=0.46\columnwidth]{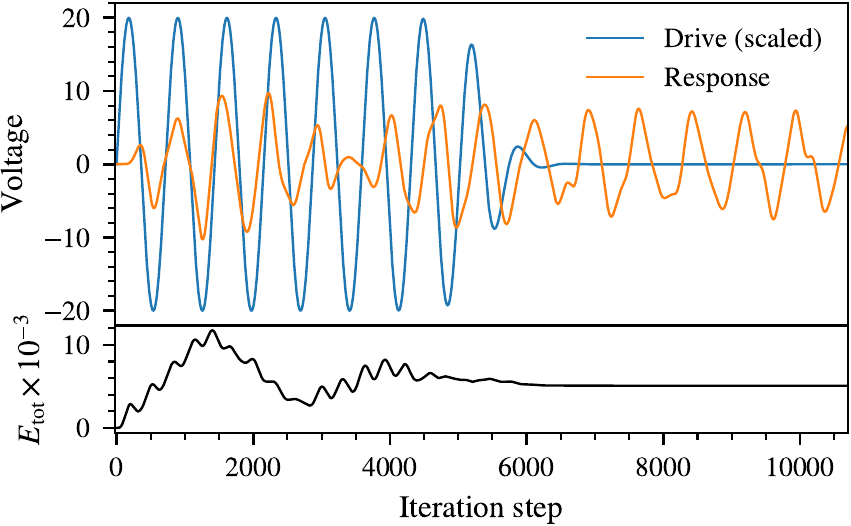}
	\caption{Driving voltage (blue) and oscillation of a node in the 4th layer (orange) over time (measured in units of Runge Kutta iteration steps). The driving frequency is set \emph{precisely at} the fundamental resonance of the system in the left panel and \emph{away} from any resonance in the right panel. The lower parts of both panels show the corresponding total energy.} 
	\label{fig:EnergyOverTime1}
\end{figure}

Due to the external driving force in the center of the tiling, an oscillation of the voltage throughout the entire lattice is initiated. In order to locate the lowest eigenfrequency of the system, we gradually increase $ \omega_d $ and detect the points $ \omega_{c,0} < \omega_{c,1} < \ldots$ where the system resonates. Most conveniently these frequencies are found by monitoring the total energy, given as 
\begin{align}
	\label{eq:total_energy}
	E_{\text{tot}}\,=\,\frac{1}{2}\,C\,\sum_{j}\,U^2_{j}\,+\,\frac{1}{2}\,L\,\sum_{k|j}\,\,I^2_{jk}.
\end{align}
Let us stress that this energy is a physically different quantity than the gravitational energy defined in \eqref{eq:Energyfunctional}. The latter is the energy of the gravitational system with respect to the Euclidean time at the boundary of EAdS$_2$, while the former is the energy of the electrical circuit obtained from Kirchhoff's laws and defined with respect to an auxiliary time parameter. While the energy in \eqref{eq:Energyfunctional} is the quantity responsible for the BF bound, the circuit energy in \eqref{eq:total_energy} is used in the simulations to monitor the response of the voltage with respect to the driving force and thus find the resonant frequencies. The quantity defined in \eqref{eq:total_energy} is expected to exhibit a pronounced peak whenever a resonance frequency is hit. In Fig.~\ref{fig:energy_spectrum} we plot the energy against the frequency of the driving force. Indeed, we observe a number of strong peaks associated to the resonance modes of the circuit. All the results shown in the figures are for the exemplary case of a \{7,3\} hyperbolic tiling.

\begin{figure}[t]
	\includegraphics[width=0.55\columnwidth]{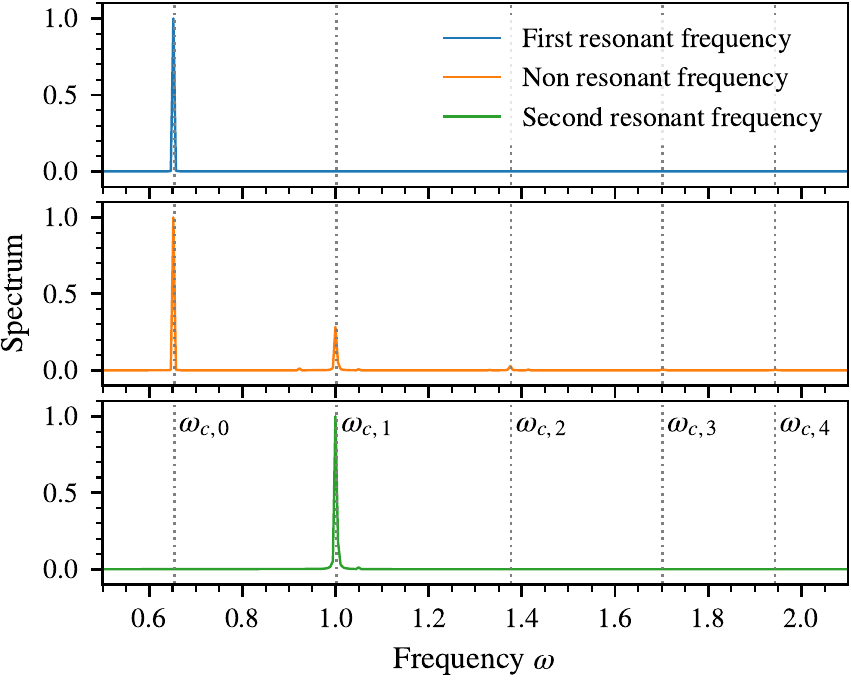}
	\caption{Examples of frequency response spectra of a driven oscillation with radial cutoff $\theta_c\approx 1.5644098$ for three different driving frequencies. In the upper panel $\omega_d=\omega_{c,0} \approx 0.653589$, in the lower panel $\omega_d=\omega_{c,1} \approx 1.002416$ and in the middle panel $\omega_d= 0.8$.} 
	\label{fig:FourierSpectrum}
\end{figure}

In order to verify that these are in fact the eigenmodes of the system, we shut down the driving force once the system has reached a state of steady oscillation. For reasons of numerical stability, this is done smoothly over the period of a few oscillations. If the eigenmode is met, we expect an unchanged steady oscillation of the system, even without the external drive, which is indeed confirmed in our simulations. Examples are shown in Fig.~\ref{fig:EnergyOverTime1}. During this phase of a ``free'' oscillation, the total energy of the system must be conserved since our circuit is assumed to be ``ideal'' and hence dissipation free. From the technical perspective conservation of energy represents an important validation for numerical stability. As can be seen in Fig.~\ref{fig:EnergyOverTime1}, obtained through the RK4 method with an error of only $\mathcal{O}((\Delta t)^4)$,  the energy indeed stays approximately constant during the further evolution of the system. Finally, as a second verification, we compute the Fourier spectra of the free oscillation, as shown in Fig.~\ref{fig:FourierSpectrum}. For the fundamental mode $ \omega = \omega_{c,0} $ (upper panel in the figure) we find only one sharp peak in an otherwise empty spectrum, as expected.

Once we have located and confirmed the fundamental eigenmode, we continue to precisely pinpoint the corresponding frequency $ \omega_{c,0} $ by fitting a Lorentzian profile to the peak in the energy-frequency diagram in Fig.~\ref{fig:energy_spectrum} and read off the maximum. This frequency can then be translated into the corresponding Klein Gordon mass squared as described in Equation~(15) in the main text.


%

\end{document}